%% file: g2_lanl.tex
\documentstyle[prl,aps,epsfig]{revtex}
\input D2DEF.tex
\input G2DEF.tex

\input ERROR_COEF_DEF.tex
\begin{document}
\renewcommand{\thefootnote}{\fnsymbol{footnote}}
\vskip -1.in
\begin{flushright}
{\small
  SLAC-PUB-8813\\
  April 19, 2002 \\
}
\end{flushright}
\begin{center}{\bf\large

Precision Measurement of the Proton and Deuteron Spin Structure Functions 
g$_{2}$ and Asymmetries A$_{2}$\footnote{Work supported by
Department of Energy contract  DE--AC03--76SF00515.}}

\vskip .1in
{
The E155 Collaboration \break  
P.~L.~Anthony,$^{12}$  
R.~G.~Arnold,$^{1,\dag\dag}$
T.~Averett,$^{15}$
H.~R.~Band,$^{16}$
N.~ Benmouna,$^{1}$ 
W.~ Boeglin,$^{5}$ 
H.~Borel,$^{4}$
P.~E.~Bosted,${^{1,\dag\dag}}$
S.~L.~B${\ddot {\rm u}}$ltmann,$^{14,\S\S}$
G.~R.~Court,$^{6}$
D.~Crabb,$^{14}$
D.~Day,$^{14}$
P.~Decowski,$^{11}$
P.~DePietro,$^{1}$
H.~Egiyan,$^{15}$
R.~Erbacher,$^{12,\infty}$
R.~Erickson,$^{12}$
R.~Fatemi,$^{14}$ 
E.~Frlez,$^{14}$
K.~A.~Griffioen,$^{15}$
C.~Harris,$^{14}$
E.~W.~Hughes,$^{2}$
C. Hyde-Wright,$^{9}$
G. Igo,$^{3}$
J. Johnson,$^{12}$
P. King,$^{15}$
K.~Kramer,$^{15}$ 
S.~E.~Kuhn,$^{9}$
D. Lawrence,$^{8}$  
Y. Liang,$^{1}$
R.~Lindgren,$^{14}$
R.~M.~Lombard-Nelsen,$^{4}$
P.~McKee,$^{14}$
D.E. McNulty,$^{14}$
W. Meyer,$^{14,\flat}$
G.~S.~Mitchell,$^{16,\S}$
J.~Mitchell,$^{13}$
M.~Olson,$^{10}$
S.~Penttila,$^{7}$`
G.~A.~Peterson,$^{8}$
R.~Pitthan,$^{12}$
D.~Pocanic,$^{14}$
R.~Prepost,$^{16}$   
C.~Prescott,$^{12}$
B.~A.~Raue,$^{5}$
D.~Reyna,$^{1,\heartsuit}$
P. Ryan,$^{15}$
L.~S.~Rochester,$^{12}$
S.~Rock,${^{1,\dag\dag}}$
O.~Rondon-Aramayo,$^{14}$
F.~Sabatie,$^{9,\spadesuit}$
T.~Smith,$^{7}$
L.~Sorrell,$^{1}$
S.~St.Lorant,$^{12}$
Z.~Szalata,$^{1,\dag}$
Y.~Terrien,$^{4}$
A.~Tobias,$^{14}$
T.~Toole,$^{1,\ddag}$
S.~Trentalange,$^{3}$
F.~R.~Wesselmann,$^{9}$
T.~R.~Wright,$^{16}$
M.~Zeier,$^{14}$
H. ~Zhu,$^{14}$
B.~Zihlmann$^{14}$
}
{\it
\baselineskip 14 pt
\vskip 0.1cm
\vskip 0.1 cm

{$^{1}$American University, Washington, D.C. 20016}  \break
{$^{2}$California Institute of Technology, Pasadena, California 91125} \break
{$^{3}$University of California, Los Angeles, California 90095} \break
{$^{4}$DAPNIA-Service de Physique` Nucleaire, CEA-Saclay,
F-91191 Gif/Yvette Cedex, France} \break
{$^{5}$Florida International University, Miami, Florida 33199.} \break
{$^{6}$University of Liverpool, Liverpool L69 3BX, United Kingdom } \break
{$^{7}$Los Alamos National Laboratory, Los Alamos, New Mexico 87545} \break
{$^{8}$University of Massachusetts, Amherst, Massachusetts 01003} \break
{$^{9}$Old Dominion University, Norfolk, Virginia 23529} \break
{$^{10}$St. Norbert College, De Pere, WI 54115} \break
{$^{11}$Smith College, Northampton, Massachusetts 01063} \break
{$^{12}$Stanford Linear Accelerator Center, Stanford, California 94309 } \break
{$^{13}$Thomas Jefferson National Accelerator Facility, Newport News, Virginia
23606} \break
{$^{14}$University of Virginia, Charlottesville, Virginia 22901} \break
{$^{15}$The College of William and Mary, Williamsburg, Virginia 23187} \break
{$^{16}$University of Wisconsin, Madison, Wisconsin 53706} \break
\break \break
{\it To be submitted to Physical Review Letters} \break}

\end{center}
 
\begin{abstract}
We have measured the spin structure functions $g_{2}^{p}$ and $g_{2}^{d}$  and 
the virtual photon asymmetries $A_{2}^{p}$ and $A_{2}^{d}$ over the 
kinematic range
$0.02\leq x \leq 0.8$ and $0.7 \leq Q^{2} \leq 20$ GeV$^2$  
by scattering 29.1 and 32.3
GeV longitudinally polarized electrons from transversely polarized NH$_3$ and 
$^6{\rm LiD}$ targets.
Our measured $g_2$  approximately follows the twist-2 Wandzura-Wilczek 
calculation.
The twist-3 reduced matrix elements  $d_{2}^{p}$  and 
$d_{2}^{n}$ are less than two standard deviations from zero. The data are
inconsistent with the Burkhardt-Cottingham sum rule if there is no pathological
behavior as $x\rightarrow 0$. The  Efremov-Leader-Teryaev integral is 
consistent with zero within our measured kinematic range.  The absolute 
value of  A$_{2}$ is significantly smaller than the $A_2<\sqrt{R(1+A_1)/2}$ 
limit. 
\end{abstract}
\vskip .2in


\input g2_body.tex

\newpage
\renewcommand{\arraystretch}{1.05}
\input APERP.tex   
\renewcommand{\arraystretch}{1.00}
\newpage
\input A2-XG2-PLB.tex  
\newpage

\begin{figure}
{\epsfig{figure=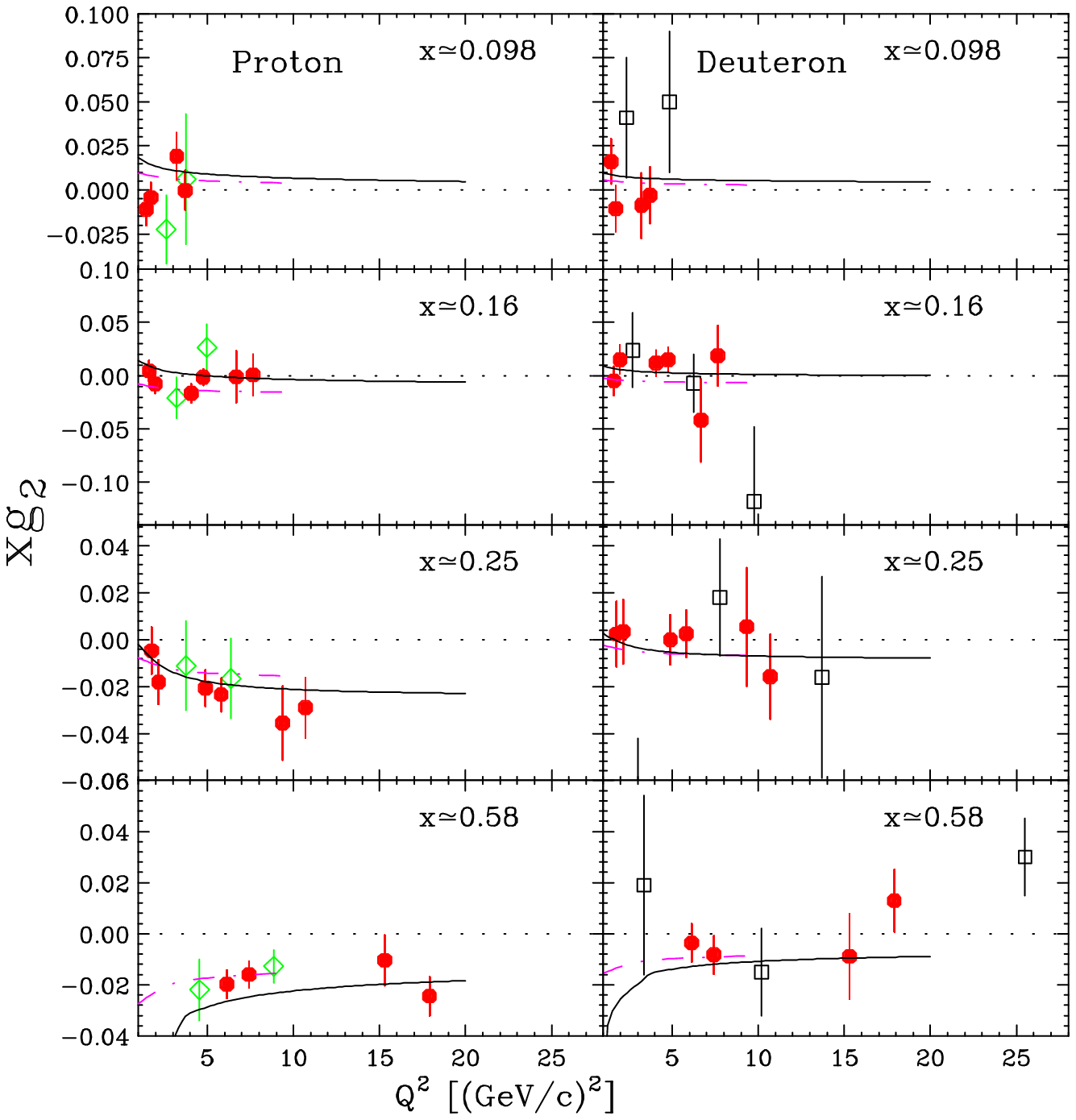,width=14.0cm}}
\vskip .2in
\caption{$xg_2$ for the proton and deuteron as a function of $Q^2$ for selected
values of $x.$  Data are for this experiment (solid),  
E143\protect\cite{E143} (open diamond) and E155\protect\cite{E155g2}
(open square). The errors are statistical; the systematic
errors are negligible. The curves show $xg_2^{WW}$ (solid) and the bag model 
calculation of Stratmann\protect\cite{Stratmann} (dash-dot).  }
\label{fg:q2}
\end{figure}

\begin{figure}
{\epsfig{figure=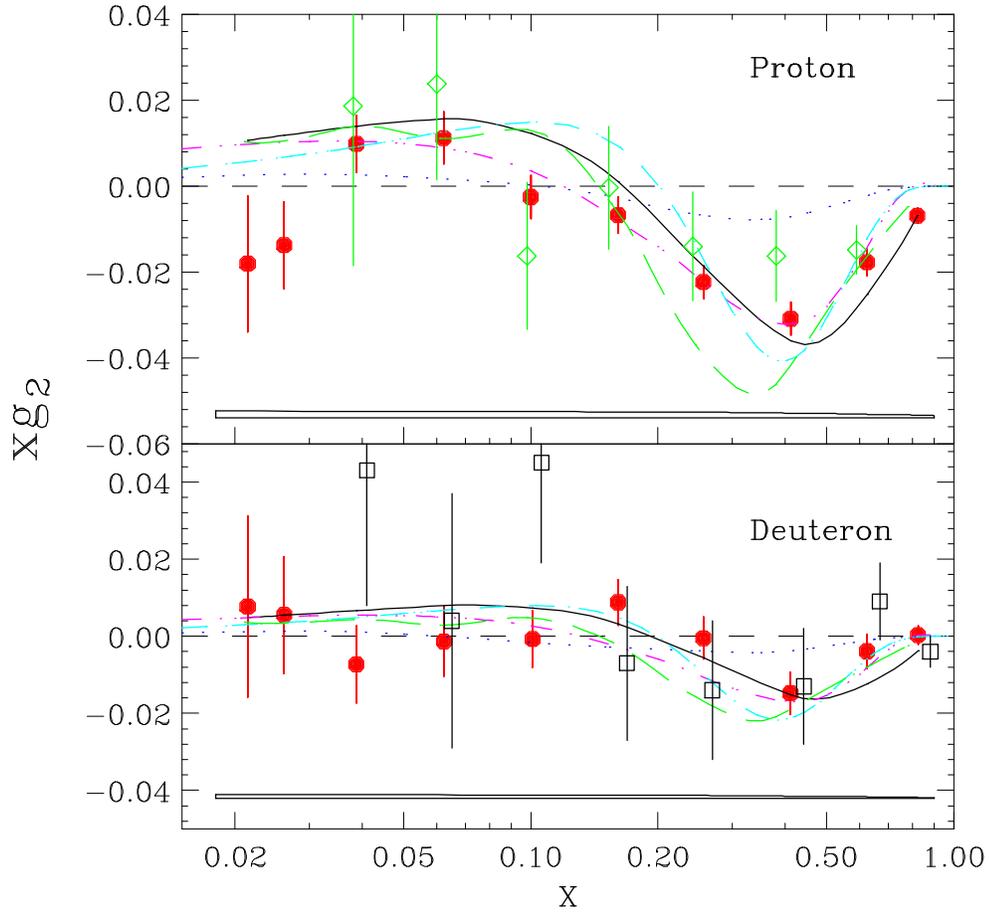}}
\vskip .2in
\caption{The structure function $x$g$_2$ for all spectrometers
combined (solid circle) and data from E143\protect\cite{E143_2} (open diamond) 
and E155\protect\cite{E155g2} (open square). 
The errors are statistical;
the systematic errors are shown at the bottom.
Also shown is our twist-2 g$_2^{WW}$ at the average $Q^2$ of this
experiment at each value of $x$ (solid line). The curves are the bag model calculations of 
Stratmann\protect\cite{Stratmann} (dash-dot) and Song\protect\cite{Song} (dot) and the 
chiral soliton models of Weigel and Gamberg\protect\cite{WGR} (short dash) and
 Wakamatsu\protect\cite{Waka} (long dash).}
\label{fg:xg2}
\end{figure}

\begin{figure}
{\epsfig{figure=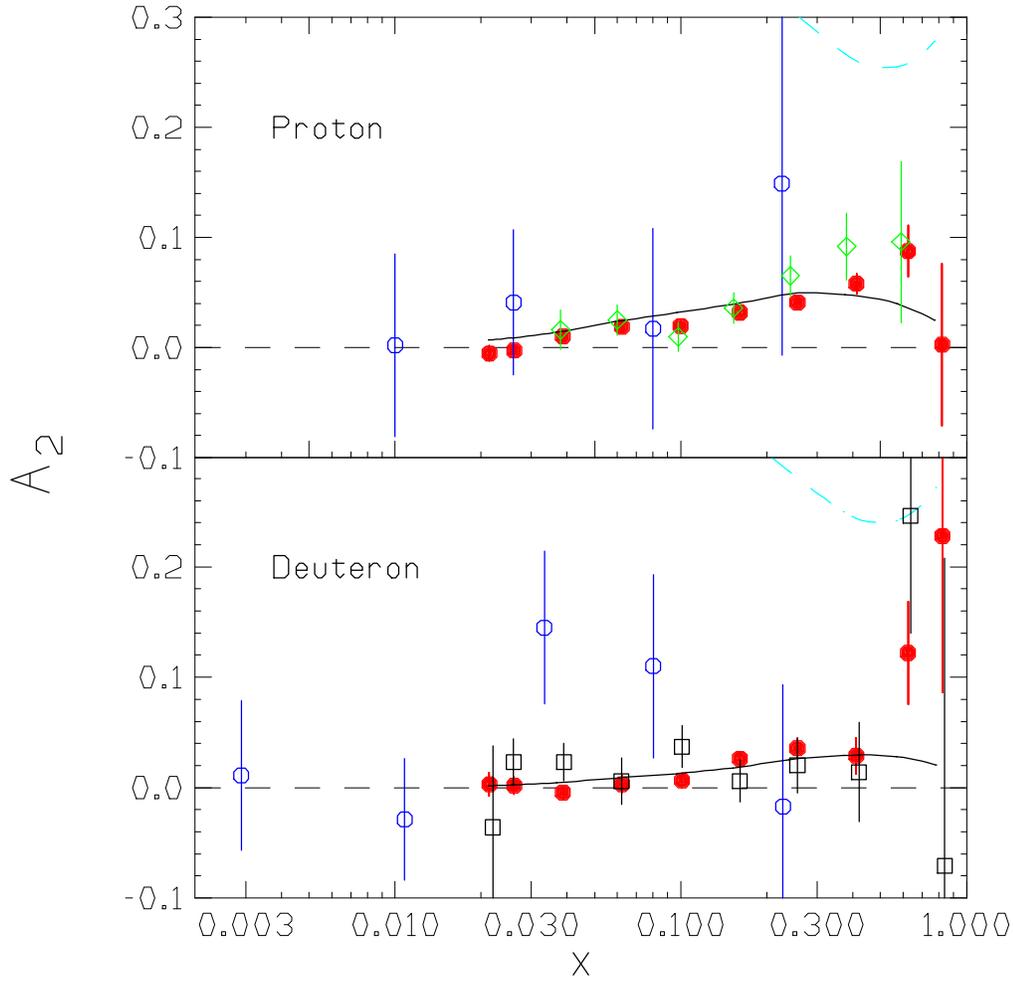}}
\vskip .2in
\caption{The asymmetry $A_2$ for all spectrometers
combined (solid circle) and data from E143\protect\cite{E143_2} (open diamond), 
E155\protect\cite{E155g2} (open square), and 
SMC\protect\cite{SMCg2} (open circles).
The errors are statistical;
the systematic errors are negligible.
Also shown is $A_2^{WW}$ calcultated from the twist-2 g$_2^{WW}$ at the average 
$Q^2$ of this experiment at each value of $x$ (solid line). The upper 
Soffer limit\protect\cite{soffer} is the dashed curve at the upper right.}
\label{fg:A2}
\end{figure}

\begin{figure}
{\epsfig{figure=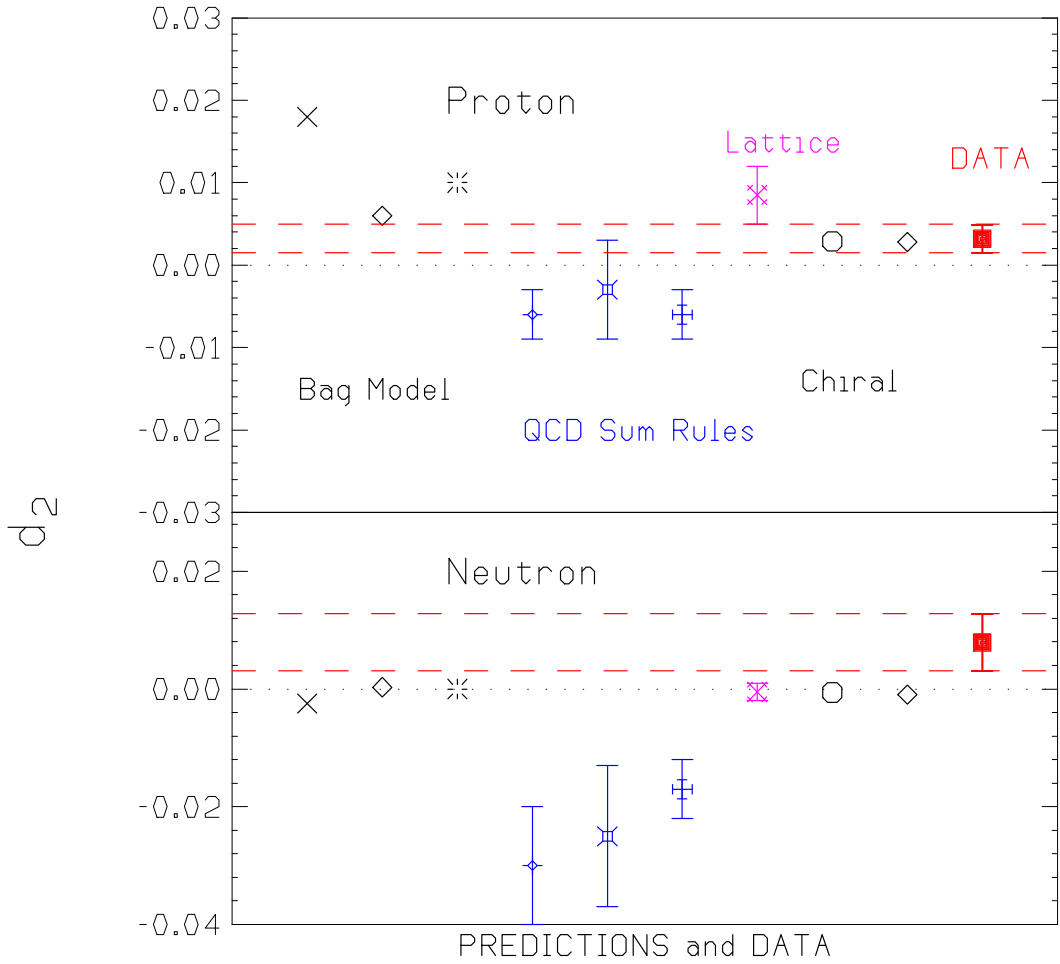,width=14.0cm}}
\vskip .2in
\caption{The twist-3 matrix element $d_2$  for the 
proton and neutron from the combined data from
SLAC experiments E142\protect\cite{E142_2}, E143\protect\cite{E143}, 
E154\protect \cite{E154_2} and E155  (Data).  Also shown are
theoretical model values  from left to right: 
Bag Models\protect\cite{Song,Stratmann,Ji}, QCD Sum Rules 
\protect\cite{Stein,BBK,Ehrnsperger}, Lattice QCD \protect\cite{LQCD} and
Chiral Soliton Models\protect\cite{WGR,Waka}.
The region between the dashed lines indicates the experimental errors.}
\label{fg:theory}
\end{figure}

\end{document}

%% file: D2DEF.tex
\def\dtpAv{$ 0.0032~$}
\def\dtnAv{$ 0.0079~$}

\def\dtpAvEr{0.0017~}
\def\dtnAvEr{0.0048~}

\def\dtNSAv{$-0.0141~$}
\def\dtNSAvEr{0.0170~}

%% file: G2DEF.tex
\def\MultSysErrP{5.1}
\def\MultSysErrD{6.2}
\def\ChiGwwd{1.2~}

\def\ChiGwwp{3.1~}
\def\NChiGwwp{10~}

\def\ChiZerop{15.5}

\def\d2Weigelp{ 0.0029~}
\def\d2WeigelN{-0.0006~}
\def\dtp{$ 0.0025~$}
\def\dtpEr{0.0016~}
\def\dtd{$ 0.0054~$}
\def\dtdEr{0.0023~}

\def\BCp{$-0.044~$}
\def\BCpEr{0.008~}
\def\BCd{$-0.008~$}
\def\BCdEr{0.012~}
\def\BCAvp{$-0.042~$}

\def\BCAvd{$-0.006~$}

\def\BCAvdErTot{0.011}
\def\BCAvpErTot{0.008}
\def\ELT{$-0.013$}
\def\ELTEr{ 0.008}
\def\ELTAv{$-0.011$}

\def\ELTAvErTot{ 0.008}
\def\xtgtp{$-0.0072~$}
\def\xtgtpErr{ 0.0005~}
\def\xtgtd{$-0.0019~$}
\def\xtgtdErr{ 0.0007~}

%% file: ERROR_COEF_DEF.tex
\def\SysErrELT{0.002~}
\def\SysErrBCP{0.003~}
\def\SysErrBCD{0.002~}
\def\SysErrDtP{0.0010~}
\def\SysErrDtD{0.0005~}
\def\SysErrxtgtP{0.0003~}
\def\SysErrxtgtD{0.0001~}
\def\SysErrCoef1d{ 0.0009~}
\def\SysErrCoef2d{-0.0009~}
\def\SysErrCoef1p{ 0.0016~}
\def\SysErrCoef2p{-0.0012~}

%% file: g2_body.tex
The deep inelastic spin structure functions of the nucleons, $g_{1}(x,Q^2)$ and 
$g_{2}(x,Q^2)$, depend on the spin distribution of the partons and their
correlations.  The function $g_1$ can be primarily understood in
terms of the quark parton model (QPM) and perturbative QCD 
with higher twist terms at low $Q^2.$   
The function $g_2$ is of particular interest since it has contributions from 
quark-gluon correlations and other higher twist terms at leading order in
$Q^2$ which cannot be described perturbatively.
By interpreting $g_{2}$ using the operator product expansion (OPE)
\cite{Vain,Jaffe}, it is possible to study contributions to
the nucleon spin structure beyond the simple QPM. 

The structure function $g_2$ can be written\cite{CPR}:
\begin{equation}
g_2(x,Q^2) = g_2^{WW}(x,Q^2) +\overline{g_2}(x,Q^2)
\end{equation}
in which
\begin{eqnarray}
g_2^{WW}(x,Q^{2})&=&-g_{1}(x,Q^{2}) + \int_{x}^{1}
\frac{g_1(y,Q^{2})}{y} dy,\nonumber \\ 
\overline{g_2}(x,Q^2) &=& -\int_x^1{\partial\over \partial y}\biggl({m\over M}h_T(y,Q^2)+\xi(y,Q^2)\biggr){dy\over y},  \nonumber
\end{eqnarray}
$x$ is the  Bjorken scaling variable and $Q^{2}$ is the absolute value
of the virtual photon four-momentum squared.
The twist-2 term $g_2^{WW}$ was derived by  Wandzura and Wilczek\cite{g2ww}
and depends only on $g_1$\cite{SMC,E143,E142_2,E154,E155,Hermes}. 
The function $h_T(x,Q^2)$ is an additional  twist-2
contribution\cite{CPR,Song} 
that depends on the transverse polarization density in the 
nucleon. The $h_T$ contribution to  $\overline{g_2}$  is  suppressed 
by the ratio of the quark to nucleon masses $m/M$\cite{Song} 
and its effect is thus  small for up and down quarks and is assumed negligible
in the theoretical models referenced in this paper. 
The twist-3 part ($\xi$) comes from quark-gluon correlations and is the main
focus of our study.

Low-precision measurements of $g_2$ and $A_2$ exist for the
proton and  deuteron~\cite{SMCg2,E143_2,E155g2}, as well as 
for the neutron\cite{E142_2,E154_2}.
In this Letter, we report new, precise measurements of $g_{2}$ and $A_{2}$
for the proton and deuteron.

Electron beams with energies of 29.1 and 32.3 GeV
and longitudinal polarizations of 
$P_b=(83.2\pm3.0)$\%  struck  approximately transversely polarized 
NH$_3$\cite{E143} (average polarization $<P_t>=0.70)$ or $^6$LiD\cite{LiD} 
$(<P_t>=0.22)$ targets.  The beam helicity
was randomly chosen pulse by pulse.
Scattered electrons were detected in
three independent spectrometers centered at 2.75$^\circ$, 5.5$^\circ$, and
10.5$^\circ$.  The two small-angle spectrometers
were the same as in SLAC E155\cite{E155}, while the large-angle spectrometer
had additional hodoscopes and a more efficient pre-radiator shower counter.
Further information on the experimental apparatus can be found in
references\cite{E143,E154,E155}. The approximately equal amounts of data  
taken with the two beam energies and opposites signs of target 
polarization gave consistent results.

The measured asymmetry, $\tilde{A_\perp}$, differs from $A_\perp$ because 
the target polarizations were not exactly perpendicular to the beam line. 
For each kinematic bin $\tilde{A_\perp}$  was formed using
\begin{eqnarray}
\tilde{A_\perp}={1\over f_{RC}}\bigg[{C_{1}\over fP_t}
\bigg( \bigg({N_L-N_R \over
N_L+N_R} \bigg) {1 \over P_b}-A_{EW} \bigg) +C_2\bigg] +A_{RC}
\end{eqnarray}
where $N_L$ and $N_R$ are the measured counting rates from the 
two beam helicities, including small corrections for 
pion and charge symmetric backgrounds, dead-time and tracking efficiency,
and $A_{EW}$ is the electroweak asymmetry ($\approx 8\times 10^{-5}Q^2$).
The target dilution factor,  $f$, 
is the fraction of free polarizable
protons ($\approx 0.13$) or deuterons ($\approx 0.18$) for a given spectrometer acceptance.
For the proton target, the nuclear correction $C_1\approx 0.98$ is due to the 
polarization of the $^{15}$N and $C_2$=0. The deuteron data were extracted
from the $^6$LiD results by applying a slightly $x$-dependent 
nuclear correction  $C_1\approx 0.52$ to  account for the lithium  and  
deuterium nuclear wave functions with $^6$Li$ \sim \alpha + d$\cite{LiD}.
An additional  correction $C_2(x)\sigma_p/\sigma_d \tilde{A^p_\perp}$,
($C_2\approx -0.042$) 
accounts for the $\sim4\%$ polarized $^7$Li in the target.
The quantities $f_{RC}$ and $A_{RC}$ are radiative corrections 
determined using a method similar to E143\cite{E143}.  
The quantity $1-f_{RC}$ was calculated as 
the proportion of events in a bin coming from elastic
and quasi-elastic tails, and $A_{RC}$ included polarization-dependent
elastic and quasi-elastic as well as inelastic and vertex corrections.
The radiative dilution factor $f_{RC}$
has the effect of increasing the statistical errors at low $x$.
Uncertainties in the
radiative corrections were estimated by varying the input models over
a range consistent with the measured data. 

The detailed results for $\tilde{A_\perp}$ are shown in Table \ref{tb:APERP}.
Because  $\tilde{A_\perp}$ is close to zero, the relative 
statistical errors are always greater than 25\%.
The uncertainties due to target and beam polarization and dilution 
factor combined are \MultSysErrP\% (proton) and \MultSysErrD\% (deuteron). 
They are multiplicative and small compared to the statistical errors. The total
systematic error is also shown in Table  \ref{tb:APERP}.

We determined $g_2(x,Q^2)$ and $A_2(x,Q^2)$ from $\tilde{A_\perp}$ 
(dominant contribution)
and the previously measured $g_1$ (small contribution) using:
\begin{eqnarray}
g_2 &=&
\frac{yF_1}{2E'(\cos\Theta -\cos\alpha)} \Bigg[\tilde{A_\perp} \nu 
\frac{(1 + \epsilon R)}{1 - \epsilon} -
\frac{g_1}{F_1} [E \cos\alpha + E' \cos\Theta]\Bigg] \\
A_2&=& \gamma(g_1 +g_2)/F_1
\end{eqnarray}
where 
$\cos\Theta=\sin\alpha\sin\theta\cos\Phi+\cos\alpha\cos\theta$, 
 $\theta$ is the spectrometer angle,
$\Phi$ is the angle between the spin plane and the scattering plane,
$\alpha=92.4^\circ$ is the angle of the target polarization  with respect to
the beam direction,  $y= \nu/E$, $\nu = E-E'$,
$E$  and $E'$ are the incident and scattered electron energies, 
$\epsilon^{-1}= 1+2\left[1+1/\gamma^2\right]{\rm{tan}}^{2}(\theta /2)$,
$\gamma=\sqrt{Q^2/\nu^2}$
and $F_1 = F_2(1 + 4M^2x^2/Q^2)/[2x(1+R)].$
We used a new $Q^2$-dependent parameterization of $g_1$ \cite{E155} 
world data,  the NMC fit to
$F_{2}(x,Q^{2})$~\cite{NMC} and the SLAC fit to
$R(x,Q^{2})=\sigma_L/\sigma_T$~\cite{R1998}. The structure functions for
p, d, and n are related by $g_2^d = (g_2^p + g_2^n)(1-1.5\omega_D)/2$, 
where $\omega_D =0.05$,
the fraction of D-wave in the deuteron wave function.  

Results for $A_2$ and  $xg_2$ for the three spectrometers and two energies
are given in Table~\ref{tb:A2-XG2} with statistical  errors. The systematic
error on $xg_2$ is much smaller than the statistical error and is 
given approximately by $a + bx$ where 
$a_p(a_d)$=$0.0016( 0.0018)$  and $b_p(b_d)$=$-0.0014(-0.0021).$ It includes 
the systematic errors on $\tilde{A_\perp}$ as well as a 5\% normalization \
uncertainty on $g_1$.
The data cover the kinematic range $0.02\leq x \leq 0.8$ and 
$0.7\leq Q^{2} \leq 20$ GeV$^{2}$ with an average $Q^2$ of $5$
GeV$^2$.  Figure \ref{fg:q2} shows the  values of $xg_2$ as a function of
$Q^{2}$ for several values of $x$ along with results from E143\cite{E143} and
E155\cite{E155g2}.
The data approximately follow the $Q^{2}$ dependence of $g_2^{WW}$
(solid curve), although for the proton, the data points are lower than 
$g_2^{WW}$  at low and intermediate $x$ and higher at high $x$.
The predictions of Stratmann\cite{Stratmann}  are closer to the 
data.

To get average values at the average $Q^2$ for each
$x$ bin we used the $Q^2$ dependence of  $g_2^{WW}$:
$g_2(Q^2_{avg}) = g_2(Q^2_{exp}) - g_2^{WW}(Q^2_{exp})+ g_2^{WW}(Q^2_{avg}).$
These averaged results for $A_2$ and  $xg_2$  are listed at
the bottom of  Table~\ref{tb:A2-XG2}.   Figure \ref{fg:xg2} shows the averaged 
$xg_2$ of this experiment along with  $xg_2^{WW}$ calculated 
using our parameterization of 
$g_1.$ The combined new data for p  disagree with  $g_2^{WW}$
with a $\chi^2/$dof of \ChiGwwp
for \NChiGwwp degrees of freedom. For d the new data agree with  $g_2^{WW}$
with a $\chi^2/$dof of \ChiGwwd for \NChiGwwp dof. 
The data for $g_2^p$ are also inconsistent with zero
($\chi^2/$dof=\ChiZerop) while $g_2^d$ differs from zero only at $x\sim 0.4.$
Also shown in Fig. \ref{fg:xg2} is the Bag Model calculation of 
Stratmann\cite{Stratmann} which is in good agreement with the data, 
a chiral soliton model calculation\cite{WGR} which is too negative
at $x\sim 0.4$ and the Bag Model calculation of Song\cite{Song} which is in
clear disagreement with the data.

The average values of $A_2(x)$, shown in Fig. \ref{fg:A2},  are consistent 
with zero at low $x$, increasing to
about 0.1 at the highest $x,$ significantly different than zero.   
$A_2^p$  is many standard deviations lower than the
Soffer limit\cite{soffer} of $|A_2|<\sqrt{R(1+A_1)/2} \approx 0.4$ 
for all values of $x$. The same is true for $A_2^d$, except at the highest
$x$ value where the error is large. 

The OPE allows us to write the hadronic matrix element in deep 
inelastic scattering in terms of a series of renormalized operators of
increasing twist\cite{Vain,Jaffe}.  
The moments of $g_{1}$ and $g_{2}$ for even $n\geq 2$ at fixed $Q^{2}$ can 
be related to twist-3 reduced matrix element, $d_n$, and higher twist
terms which are suppressed by powers of $1/Q$. Neglecting quark mass 
terms we find that:
\begin{equation}
d_n = 2\int_0^1dx\  x^n \biggl[ {n+1 \over n} g_2(x,Q^2) + g_1(x,Q^{2})\biggr] = 
2{n+1\over n}\int_0^1dx\  x^n \overline{g_2}(x,Q^2).
\label{eq:dn}
\end{equation}
This relation is true to all orders of $\alpha_s$\cite{JiOsborne}.
The matrix element $d_n$ measures deviations of $g_2$ from the twist-2 
$g_2^{WW}$ term.
Note that some authors\cite{Jaffe,LQCD} 
define $d_n$ with an additional factor of two.
We calculated $d_n$ using  
$d_2=3\int_0^1dx\  x^2 \overline{g_2}(x,Q^2)$ (see eq.\ref{eq:dn})
with the  assumption
that  $\overline{g_2}$ is independent of $Q^2$ in the measured region.
This is not unreasonable since $d_n$ depends only logarithmically
on $Q^2$\cite{Vain}.  The part of the integral for $x$ below the measured
region was assumed to be zero because of the $x^2$ suppression. 
For $x\geq 0.8$ we used $\overline{g_2}\propto (1-x)^m$ where $m$=2 or 3,
normalized to the data for $x\geq 0.5$. 
Because $\overline{g_2}$ is small at high $x$, the contribution was negligible
for both cases. We obtained values of
$d_{2}^p=$\dtp$\pm$\dtpEr$\pm$\SysErrDtP                
and $d_{2}^d=$\dtd$\pm$\dtdEr$\pm$\SysErrDtD         
at an average $Q^{2}$ of $5$ GeV$^{2}$.
We combined these results with those from SLAC experiments on the neutron 
(E142\cite{E142_2} and E154\cite{E154_2}) and proton and deuteron 
(E143\cite{E143} and E155\cite{E155g2})  to obtained average values 
$d_2^p=$\dtpAv$\pm$\dtpAvEr     
and $d_2^n=$\dtnAv$\pm$\dtnAvEr. 
These are consistent with zero (no twist-3) to within 2 standard deviations.
The values of the 2$^{nd}$ moments alone are: 
$\int_0^1dx\  x^2 g_2(x,Q^2)$=\xtgtp$\pm$\xtgtpErr$\pm$\SysErrxtgtP (p) and 
\xtgtd $\pm$\xtgtdErr$\pm$\SysErrxtgtD (d).

Figure \ref{fg:theory}~ shows the experimental values of $d_2$ for
proton and neutron with their error, plotted along with theoretical models 
from  left to right:
Bag Models (Song\cite{Song}, Stratmann\cite{Stratmann},
and Ji\cite{Ji}); sum rules
(Stein\cite{Stein}, BBK\cite{BBK}, Ehrnsperger\cite{Ehrnsperger});
chiral soliton models\cite{WGR,Waka};
and lattice QCD calculations ($Q^2=5$ GeV$^2$, $\beta=6.4$)\cite{LQCD}.  
The lattice and chiral
calculations are in good agreement with the proton data and two
standard deviations below the neutron data. The sum rule calculations are
significantly lower than the data. The Non Singlet $3\cdot(d_2^p -d_2^n) =$
\dtNSAv$\pm$\dtNSAvEr is consistent with an instanton vacuum 
calculation of $\sim0.001$ \cite{Weiss}.

 The Burkhardt-Cottingham sum rule\cite{buco} for $g_2$ at large $Q^2$, 
$\int_0^1 g_2(x)dx=0,$
was derived from virtual Compton scattering dispersion relations. It
does not follow from the OPE since $n=0$.
Its validity depends on the lack of singularities for $g_2$ at 
$x=0,$ and a dramatic rise of
$g_2$ at low $x$ could invalidate the sum rule\cite{Ivanov}. 
We evaluated the Burkhardt-Cottingham integral in the measured region
of $0.02\leq x \leq 0.8$ at $Q^2=5~ {\rm GeV^2}.$ 
The results for the proton and deuteron are \BCp$\pm$\BCpEr$\pm$\SysErrBCP  
and \BCd$\pm$\BCdEr$\pm$\SysErrBCD 
respectively.  Averaging with the E143 and E155 results which cover a slightly
more restrictive $x$ range gives \BCAvp$\pm$\BCAvpErTot~ 
and \BCAvd$\pm$\BCAvdErTot.    
This does not represent a conclusive test of the sum rule
because the behavior of $g_2$ as $x\rightarrow 0$ is not known. However,
if we assume that $g_2 = g_2^{WW}$ for $x<0.02$, and use the relation
$\int_0^x g_2^{WW}(y) dy = x \left[ g_2^{WW}(x) + g_1(x) \right]$,
there is an additional contribution of 0.020 (0.004) for the proton (deuteron).

The  Efremov-Leader-Teryaev (ELT) sum rule\cite{ELT} involves the valence 
quark contributions to $g_1$ and $g_2$:
$\int_0^1 x[g_1^V(x) + 2g_2^V(x)]dx=0.$
Assuming that the sea quarks are the same in protons and neutrons, the sum
rule takes a form
$\int_0^1 x[g_1^p(x) + 2g_2^p(x) - g_1^n(x) - 2g_2^n(x) ]dx=0.$ 
We evaluated this ELT integral in the measured region  using our $g_2$ data  
and the fit to $g_1$.
The result at $Q^2=5~{\rm GeV^2}$ is \ELT$\pm$\ELTEr$\pm$\SysErrELT, 
which is consistent with the
expected value of zero. Including the data of E143 \cite{E143} and 
E155\cite{E155g2} leads to
\ELTAv$\pm$\ELTAvErTot. The extrapolation to $x$=0 is
not known, but is suppressed by a factor of $x.$  

In summary, our results for $g_{2}$ follow approximately
the twist-2 $g_{2}^{WW}$ shape, but deviate significantly at some values of $x$.
The values obtained for the
twist-3 matrix element $d_2$ from this measurement and the SLAC 
average are less than two standard deviations from zero. The data over the
measured range are 
inconsistent with the Burkhardt-Cottingham sum rule if there is no pathological
behavior as $x\rightarrow 0$. The  ELT integral is consistent with zero within
our measured kinematic range. 

We wish to thank the personnel of the SLAC accelerator and Experimental Facilities 
Departments for
their efforts which resulted in the successful completion of the E155X
experiment.  This work was supported by the
Department of Energy; the National Science Foundation;  and the Centre
National de la Recherche Scientifique and the Commissariat a l'Energie
Atomique (French groups).\\

%% file: APERP.tex

 \begin{table} 
 \caption{ Results for $\tilde{A_\perp}$  with statistical and systematic errors for proton  and deuteron 
 at the measured $x$ and $Q^2$ [(GeV/c)$^2$]. Also 
 shown are the additive and multiplicative radiative 
 corrections } 
\renewcommand{\baselinestretch}{.75}
\small\normalsize                   
\label{tb:APERP}
 \begin{tabular}{rrrrrrrr}
$x$ & $<Q^2>$ & $\tilde{A_{\perp}}^p$ & $f_{RC}^p$ & $A_{RC}^p$ & $\tilde{A_\perp}^d$ &                        $f_{RC}^d$ & $A_{RC}^d$  \\
 \hline
\multicolumn{6}{c}{$\theta\approx 2.75^\circ$; E=29.1 GeV   } \\
\hline
 0.017 &  0.63 & $  8.611 \pm4.785 \pm$0.456 &  0.811 & -0.001 & $  5.545 \pm4.476 \pm$NaN   &  0.877 &  0.000 \\
 0.020 &  0.71 & $  0.025 \pm0.015 \pm$0.001 &  0.841 & -0.001 & $  0.018 \pm0.022 \pm$0.001 &  0.896 &  0.000 \\
 0.022 &  0.76 & $  0.004 \pm0.007 \pm$0.000 &  0.861 & -0.001 & $ -0.006 \pm0.010 \pm$0.001 &  0.909 &  0.000 \\
 0.024 &  0.82 & $  0.000 \pm0.005 \pm$0.000 &  0.880 & -0.001 & $ -0.003 \pm0.008 \pm$0.000 &  0.922 &  0.000 \\
 0.027 &  0.86 & $  0.008 \pm0.005 \pm$0.000 &  0.904 & -0.001 & $ -0.007 \pm0.008 \pm$0.001 &  0.937 &  0.000 \\
 0.031 &  0.91 & $  0.001 \pm0.005 \pm$0.000 &  0.921 & -0.001 & $  0.006 \pm0.008 \pm$0.000 &  0.948 &  0.000 \\
 0.035 &  0.96 & $ -0.002 \pm0.005 \pm$0.000 &  0.935 & -0.001 & $  0.008 \pm0.008 \pm$0.001 &  0.957 &  0.000 \\
 0.039 &  1.02 & $ -0.011 \pm0.005 \pm$0.001 &  0.946 & -0.001 & $  0.000 \pm0.008 \pm$0.000 &  0.957 &  0.000 \\
 0.044 &  1.07 & $ -0.015 \pm0.005 \pm$0.001 &  0.954 & -0.001 & $  0.008 \pm0.007 \pm$0.001 &  0.970 &  0.000 \\
 0.049 &  1.13 & $ -0.008 \pm0.005 \pm$0.000 &  0.962 & -0.001 & $ -0.003 \pm0.007 \pm$0.000 &  0.975 & -0.001 \\
 0.056 &  1.18 & $ -0.004 \pm0.005 \pm$0.000 &  0.967 & -0.001 & $  0.009 \pm0.007 \pm$0.001 &  0.978 & -0.001 \\
 0.063 &  1.23 & $ -0.005 \pm0.005 \pm$0.000 &  0.971 & -0.002 & $  0.008 \pm0.007 \pm$0.001 &  0.981 & -0.001 \\
 0.071 &  1.29 & $ -0.010 \pm0.005 \pm$0.001 &  0.975 & -0.002 & $  0.010 \pm0.007 \pm$0.001 &  0.984 & -0.001 \\
 0.079 &  1.34 & $ -0.002 \pm0.005 \pm$0.000 &  0.979 & -0.002 & $ -0.004 \pm0.007 \pm$0.000 &  0.986 & -0.001 \\
 0.089 &  1.40 & $ -0.004 \pm0.005 \pm$0.000 &  0.981 & -0.002 & $ -0.006 \pm0.007 \pm$0.000 &  0.987 & -0.001 \\
 0.101 &  1.45 & $  0.006 \pm0.005 \pm$0.001 &  0.982 & -0.002 & $  0.007 \pm0.007 \pm$0.001 &  0.989 & -0.001 \\
 0.113 &  1.50 & $  0.002 \pm0.005 \pm$0.001 &  0.986 & -0.002 & $ -0.019 \pm0.007 \pm$0.001 &  0.990 & -0.001 \\
 0.128 &  1.54 & $ -0.005 \pm0.005 \pm$0.001 &  0.985 & -0.002 & $  0.002 \pm0.008 \pm$0.000 &  0.991 & -0.001 \\
 0.144 &  1.59 & $ -0.007 \pm0.005 \pm$0.001 &  0.986 & -0.002 & $  0.004 \pm0.008 \pm$0.000 &  0.991 & -0.001 \\
 0.162 &  1.63 & $ -0.003 \pm0.005 \pm$0.000 &  0.987 & -0.002 & $  0.002 \pm0.008 \pm$0.000 &  0.991 & -0.001 \\
 0.182 &  1.67 & $ -0.002 \pm0.005 \pm$0.000 &  0.987 & -0.002 & $ -0.007 \pm0.008 \pm$0.000 &  0.991 & -0.001 \\
 0.205 &  1.71 & $  0.009 \pm0.005 \pm$0.001 &  0.988 & -0.002 & $  0.000 \pm0.009 \pm$0.000 &  0.991 & -0.001 \\
 0.230 &  1.74 & $ -0.008 \pm0.006 \pm$0.001 &  0.987 & -0.002 & $ -0.003 \pm0.009 \pm$0.000 &  0.991 & -0.001 \\
 0.259 &  1.77 & $  0.001 \pm0.006 \pm$0.000 &  0.987 & -0.002 & $ -0.009 \pm0.009 \pm$0.001 &  0.991 & -0.001 \\
 0.292 &  1.80 & $ -0.011 \pm0.006 \pm$0.001 &  0.986 & -0.002 & $  0.003 \pm0.010 \pm$0.000 &  0.990 & -0.001 \\
 0.329 &  1.83 & $ -0.009 \pm0.006 \pm$0.001 &  0.983 & -0.002 & $  0.009 \pm0.010 \pm$0.001 &  0.989 & -0.001 \\
 \hline
\multicolumn{6}{c}{$\theta\approx 5.5^\circ$; E=29.1 GeV    } \\
\hline
 0.065 &  2.32 & $ -0.013 \pm0.053 \pm$0.001 &  0.951 &  0.002 & $ -0.023 \pm0.076 \pm$0.001 &  0.965 &  0.001 \\
 0.071 &  2.49 & $  0.018 \pm0.022 \pm$0.001 &  0.958 &  0.003 & $  0.042 \pm0.032 \pm$0.003 &  0.970 &  0.001 \\
 0.080 &  2.71 & $  0.010 \pm0.014 \pm$0.001 &  0.967 &  0.003 & $  0.049 \pm0.021 \pm$0.003 &  0.977 &  0.001 \\
 0.090 &  2.94 & $ -0.011 \pm0.011 \pm$0.001 &  0.975 &  0.003 & $ -0.008 \pm0.016 \pm$0.001 &  0.982 &  0.002 \\
 0.101 &  3.17 & $ -0.004 \pm0.009 \pm$0.000 &  0.981 &  0.003 & $ -0.010 \pm0.014 \pm$0.001 &  0.986 &  0.002 \\
 0.114 &  3.40 & $  0.021 \pm0.008 \pm$0.001 &  0.986 &  0.003 & $ -0.013 \pm0.013 \pm$0.001 &  0.989 &  0.002 \\
 0.128 &  3.62 & $  0.011 \pm0.008 \pm$0.001 &  0.989 &  0.003 & $  0.000 \pm0.012 \pm$0.000 &  0.992 &  0.002 \\
 0.144 &  3.85 & $ -0.017 \pm0.008 \pm$0.001 &  0.991 &  0.003 & $  0.004 \pm0.012 \pm$0.000 &  0.994 &  0.002 \\
 0.162 &  4.08 & $ -0.016 \pm0.007 \pm$0.001 &  0.994 &  0.003 & $  0.000 \pm0.011 \pm$0.000 &  0.995 &  0.002 \\
 0.182 &  4.31 & $ -0.009 \pm0.007 \pm$0.001 &  0.995 &  0.003 & $  0.017 \pm0.012 \pm$0.001 &  0.996 &  0.002 \\
 0.205 &  4.53 & $ -0.010 \pm0.008 \pm$0.001 &  0.996 &  0.003 & $  0.021 \pm0.012 \pm$0.001 &  0.997 &  0.002 \\
 0.231 &  4.74 & $ -0.020 \pm0.008 \pm$0.001 &  0.997 &  0.003 & $ -0.002 \pm0.013 \pm$0.000 &  0.998 &  0.002 \\
 0.259 &  4.94 & $  0.004 \pm0.009 \pm$0.001 &  0.998 &  0.003 & $  0.001 \pm0.014 \pm$0.001 &  0.998 &  0.002 \\
 0.292 &  5.14 & $ -0.018 \pm0.009 \pm$0.001 &  0.998 &  0.003 & $ -0.027 \pm0.015 \pm$0.002 &  0.999 &  0.001 \\
 0.328 &  5.32 & $ -0.018 \pm0.010 \pm$0.001 &  0.998 &  0.002 & $  0.008 \pm0.016 \pm$0.001 &  0.999 &  0.001 \\
 0.370 &  5.50 & $ -0.048 \pm0.011 \pm$0.002 &  0.998 &  0.002 & $  0.011 \pm0.018 \pm$0.001 &  0.999 &  0.001 \\
 0.416 &  5.66 & $ -0.025 \pm0.012 \pm$0.001 &  0.999 &  0.002 & $ -0.034 \pm0.020 \pm$0.002 &  0.999 &  0.001 \\
 0.468 &  5.82 & $ -0.049 \pm0.013 \pm$0.003 &  0.998 &  0.002 & $  0.020 \pm0.023 \pm$0.001 &  0.999 &  0.002 \\
 0.527 &  5.96 & $ -0.034 \pm0.015 \pm$0.002 &  0.998 &  0.003 & $  0.003 \pm0.027 \pm$0.000 &  0.998 &  0.002 \\
 0.592 &  6.09 & $ -0.027 \pm0.018 \pm$0.001 &  0.997 &  0.003 & $ -0.012 \pm0.033 \pm$0.001 &  0.998 &  0.003 \\
 0.667 &  6.20 & $ -0.050 \pm0.022 \pm$0.003 &  0.995 &  0.001 & $ -0.029 \pm0.041 \pm$0.002 &  0.996 &  0.001 \\
 0.750 &  6.31 & $ -0.089 \pm0.029 \pm$0.005 &  0.990 &  0.001 & $  0.025 \pm0.054 \pm$0.002 &  0.992 &  0.002 \\
 0.844 &  6.40 & $ -0.088 \pm0.047 \pm$0.004 &  0.961 &  0.004 & $  0.057 \pm0.076 \pm$0.004 &  0.972 &  0.004 \\
 \hline
\multicolumn{6}{c}{$\theta\approx 10.5^\circ$; E=29.1 GeV   } \\
\hline
 0.129 &  5.36 & $ -0.052 \pm0.031 \pm$0.003 &  0.963 & -0.005 & $ -0.028 \pm0.053 \pm$0.002 &  0.972 & -0.002 \\
 0.144 &  5.89 & $ -0.071 \pm0.025 \pm$0.004 &  0.970 & -0.005 & $  0.040 \pm0.045 \pm$0.003 &  0.977 & -0.002 \\
 0.162 &  6.49 & $  0.006 \pm0.022 \pm$0.001 &  0.977 & -0.005 & $ -0.015 \pm0.039 \pm$0.001 &  0.983 & -0.002 \\
 0.182 &  7.17 & $  0.012 \pm0.020 \pm$0.001 &  0.981 & -0.005 & $  0.045 \pm0.036 \pm$0.003 &  0.986 & -0.003 \\
 0.205 &  7.91 & $ -0.021 \pm0.020 \pm$0.001 &  0.986 & -0.005 & $  0.007 \pm0.036 \pm$0.001 &  0.989 & -0.003 \\
 0.231 &  8.67 & $  0.020 \pm0.021 \pm$0.001 &  0.989 & -0.005 & $  0.005 \pm0.039 \pm$0.001 &  0.992 & -0.002 \\
 0.259 &  9.39 & $ -0.025 \pm0.022 \pm$0.002 &  0.992 & -0.005 & $ -0.044 \pm0.043 \pm$0.003 &  0.993 & -0.002 \\
 0.292 & 10.18 & $  0.029 \pm0.024 \pm$0.002 &  0.993 & -0.004 & $ -0.040 \pm0.047 \pm$0.003 &  0.995 & -0.002 \\
 0.328 & 10.96 & $ -0.028 \pm0.026 \pm$0.002 &  0.995 & -0.004 & $  0.069 \pm0.053 \pm$NaN   &  0.996 & -0.002 \\
 0.369 & 11.78 & $ -0.014 \pm0.029 \pm$0.002 &  0.997 & -0.004 & $ -0.088 \pm0.061 \pm$0.006 &  0.997 & -0.002 \\
 0.415 & 12.57 & $  0.004 \pm0.034 \pm$0.002 &  0.997 & -0.003 & $ -0.027 \pm0.072 \pm$0.003 &  0.997 & -0.002 \\
 0.467 & 13.36 & $  0.067 \pm0.041 \pm$NaN   &  0.998 & -0.003 & $ -0.014 \pm0.088 \pm$0.003 &  0.998 & -0.002 \\
 0.526 & 14.19 & $  0.032 \pm0.051 \pm$0.002 &  0.998 & -0.002 & $ -0.026 \pm0.114 \pm$0.002 &  0.999 & -0.002 \\
 0.592 & 15.02 & $  0.052 \pm0.067 \pm$0.003 &  0.999 & -0.002 & $  0.137 \pm0.158 \pm$0.008 &  0.999 & -0.002 \\
 0.666 & 15.77 & $ -0.195 \pm0.094 \pm$0.010 &  0.999 & -0.002 & $ -0.045 \pm0.229 \pm$0.003 &  0.999 & -0.002 \\
 0.749 & 16.48 & $  0.114 \pm0.142 \pm$0.006 &  0.999 & -0.002 & $ -0.035 \pm0.347 \pm$0.002 &  0.999 & -0.002 \\
 0.845 & 17.10 & $  0.299 \pm0.259 \pm$0.015 &  0.997 & -0.002 & $ -1.237 \pm0.588 \pm$0.077 &  0.998 & -0.003 \\
 \hline
\multicolumn{6}{c}{$\theta\approx 2.75^\circ$; E=32.3 GeV   } \\
\hline
 0.018 &  0.75 & $  0.071 \pm0.081 \pm$0.004 &  0.820 & -0.001 & $ -0.047 \pm0.110 \pm$0.003 &  0.882 &  0.000 \\
 0.020 &  0.81 & $ -0.002 \pm0.009 \pm$0.000 &  0.835 & -0.001 & $ -0.010 \pm0.014 \pm$0.001 &  0.892 &  0.000 \\
 0.022 &  0.87 & $  0.000 \pm0.005 \pm$0.000 &  0.861 & -0.001 & $  0.002 \pm0.008 \pm$0.000 &  0.908 &  0.000 \\
 0.024 &  0.93 & $ -0.006 \pm0.005 \pm$0.000 &  0.883 & -0.001 & $  0.009 \pm0.008 \pm$0.001 &  0.925 &  0.000 \\
 0.027 &  0.99 & $  0.003 \pm0.004 \pm$0.000 &  0.909 & -0.001 & $ -0.003 \pm0.008 \pm$0.000 &  0.939 &  0.000 \\
 0.031 &  1.05 & $  0.000 \pm0.004 \pm$0.000 &  0.925 & -0.001 & $  0.007 \pm0.008 \pm$0.000 &  0.950 &  0.000 \\
 0.035 &  1.12 & $ -0.004 \pm0.004 \pm$0.000 &  0.938 & -0.001 & $ -0.001 \pm0.008 \pm$0.000 &  0.959 &  0.000 \\
 0.039 &  1.19 & $ -0.001 \pm0.004 \pm$0.000 &  0.938 & -0.001 & $ -0.001 \pm0.007 \pm$0.000 &  0.967 &  0.000 \\
 0.044 &  1.25 & $ -0.004 \pm0.004 \pm$0.000 &  0.959 & -0.001 & $ -0.011 \pm0.007 \pm$0.001 &  0.973 &  0.000 \\
 0.049 &  1.32 & $  0.000 \pm0.004 \pm$0.000 &  0.966 & -0.001 & $ -0.008 \pm0.007 \pm$0.000 &  0.977 &  0.000 \\
 0.056 &  1.39 & $ -0.003 \pm0.004 \pm$0.000 &  0.968 & -0.002 & $  0.000 \pm0.007 \pm$0.000 &  0.981 & -0.001 \\
 0.063 &  1.46 & $ -0.008 \pm0.004 \pm$0.000 &  0.977 & -0.002 & $ -0.002 \pm0.007 \pm$0.000 &  0.983 & -0.001 \\
 0.071 &  1.54 & $ -0.006 \pm0.004 \pm$0.000 &  0.980 & -0.002 & $ -0.009 \pm0.007 \pm$0.001 &  0.987 & -0.001 \\
 0.079 &  1.61 & $  0.004 \pm0.004 \pm$0.000 &  0.982 & -0.002 & $ -0.002 \pm0.007 \pm$0.000 &  0.988 & -0.001 \\
 0.089 &  1.68 & $ -0.003 \pm0.004 \pm$0.000 &  0.985 & -0.002 & $  0.011 \pm0.007 \pm$0.001 &  0.990 & -0.001 \\
 0.101 &  1.74 & $ -0.005 \pm0.004 \pm$0.000 &  0.988 & -0.002 & $  0.001 \pm0.007 \pm$0.000 &  0.991 & -0.001 \\
 0.113 &  1.81 & $ -0.003 \pm0.005 \pm$0.000 &  0.988 & -0.002 & $  0.000 \pm0.007 \pm$0.000 &  0.992 & -0.001 \\
 0.128 &  1.87 & $ -0.003 \pm0.005 \pm$0.000 &  0.989 & -0.002 & $ -0.003 \pm0.008 \pm$0.000 &  0.993 & -0.001 \\
 0.144 &  1.93 & $ -0.002 \pm0.005 \pm$0.000 &  0.991 & -0.002 & $ -0.014 \pm0.008 \pm$0.001 &  0.994 & -0.001 \\
 0.162 &  1.98 & $ -0.007 \pm0.005 \pm$0.001 &  0.992 & -0.002 & $ -0.001 \pm0.008 \pm$0.000 &  0.994 & -0.001 \\
 0.182 &  2.03 & $  0.010 \pm0.005 \pm$0.001 &  0.993 & -0.002 & $ -0.004 \pm0.008 \pm$0.000 &  0.994 & -0.001 \\
 0.205 &  2.08 & $ -0.003 \pm0.005 \pm$0.000 &  0.992 & -0.002 & $ -0.008 \pm0.009 \pm$0.001 &  0.994 & -0.001 \\
 0.230 &  2.13 & $  0.004 \pm0.005 \pm$0.000 &  0.992 & -0.002 & $ -0.001 \pm0.009 \pm$0.000 &  0.994 & -0.001 \\
 0.259 &  2.17 & $  0.007 \pm0.005 \pm$0.000 &  0.992 & -0.002 & $ -0.008 \pm0.009 \pm$0.001 &  0.994 & -0.001 \\
 0.292 &  2.21 & $  0.002 \pm0.006 \pm$0.000 &  0.992 & -0.002 & $  0.009 \pm0.010 \pm$0.001 &  0.994 & -0.001 \\
 0.329 &  2.26 & $ -0.004 \pm0.006 \pm$0.000 &  0.990 & -0.002 & $  0.001 \pm0.010 \pm$0.000 &  0.993 & -0.001 \\
 \hline
\multicolumn{6}{c}{$\theta\approx 5.5^\circ$; E=32.3 GeV    } \\
\hline
 0.059 &  2.48 & $ -0.685 \pm0.248 \pm$0.035 &  0.940 &  0.002 & $  0.205 \pm0.369 \pm$0.013 &  0.957 &  0.001 \\
 0.064 &  2.65 & $  0.008 \pm0.029 \pm$0.001 &  0.949 &  0.002 & $  0.022 \pm0.043 \pm$0.001 &  0.963 &  0.001 \\
 0.071 &  2.87 & $ -0.019 \pm0.015 \pm$0.001 &  0.959 &  0.003 & $  0.026 \pm0.023 \pm$0.002 &  0.969 &  0.001 \\
 0.080 &  3.13 & $  0.004 \pm0.011 \pm$0.001 &  0.968 &  0.003 & $  0.016 \pm0.017 \pm$0.001 &  0.976 &  0.001 \\
 0.090 &  3.41 & $  0.004 \pm0.009 \pm$0.000 &  0.974 &  0.003 & $ -0.005 \pm0.013 \pm$0.000 &  0.981 &  0.002 \\
 0.101 &  3.67 & $ -0.012 \pm0.008 \pm$0.001 &  0.981 &  0.003 & $ -0.005 \pm0.012 \pm$0.000 &  0.986 &  0.002 \\
 0.113 &  3.95 & $  0.000 \pm0.007 \pm$0.001 &  0.985 &  0.003 & $ -0.004 \pm0.011 \pm$0.000 &  0.989 &  0.002 \\
 0.128 &  4.22 & $ -0.002 \pm0.007 \pm$0.001 &  0.991 &  0.003 & $  0.011 \pm0.010 \pm$0.001 &  0.992 &  0.002 \\
 0.144 &  4.51 & $ -0.005 \pm0.007 \pm$0.001 &  0.992 &  0.003 & $  0.014 \pm0.010 \pm$0.001 &  0.994 &  0.002 \\
 0.162 &  4.79 & $  0.002 \pm0.007 \pm$0.001 &  0.994 &  0.003 & $  0.000 \pm0.010 \pm$0.000 &  0.995 &  0.002 \\
 0.182 &  5.07 & $ -0.005 \pm0.007 \pm$0.001 &  0.995 &  0.003 & $  0.002 \pm0.011 \pm$0.000 &  0.997 &  0.002 \\
 0.205 &  5.34 & $ -0.003 \pm0.007 \pm$0.001 &  0.997 &  0.003 & $ -0.001 \pm0.011 \pm$0.000 &  0.997 &  0.002 \\
 0.230 &  5.60 & $ -0.023 \pm0.007 \pm$0.001 &  0.997 &  0.003 & $ -0.006 \pm0.012 \pm$0.001 &  0.998 &  0.002 \\
 0.259 &  5.86 & $ -0.007 \pm0.008 \pm$0.001 &  0.998 &  0.003 & $ -0.001 \pm0.013 \pm$0.001 &  0.998 &  0.001 \\
 0.292 &  6.11 & $ -0.022 \pm0.008 \pm$0.001 &  0.998 &  0.002 & $  0.016 \pm0.014 \pm$0.001 &  0.999 &  0.001 \\
 0.328 &  6.34 & $ -0.008 \pm0.009 \pm$0.001 &  0.999 &  0.002 & $ -0.034 \pm0.015 \pm$0.002 &  0.999 &  0.001 \\
 0.370 &  6.57 & $ -0.034 \pm0.010 \pm$0.002 &  0.999 &  0.002 & $ -0.025 \pm0.017 \pm$0.002 &  0.999 &  0.001 \\
 0.416 &  6.79 & $ -0.022 \pm0.011 \pm$0.001 &  0.999 &  0.002 & $ -0.023 \pm0.019 \pm$0.002 &  0.999 &  0.001 \\
 0.468 &  6.99 & $ -0.037 \pm0.012 \pm$0.002 &  0.999 &  0.002 & $ -0.001 \pm0.022 \pm$0.001 &  0.999 &  0.001 \\
 0.527 &  7.19 & $ -0.030 \pm0.014 \pm$0.002 &  0.999 &  0.002 & $ -0.014 \pm0.026 \pm$0.001 &  0.999 &  0.002 \\
 0.592 &  7.36 & $ -0.032 \pm0.016 \pm$0.002 &  0.998 &  0.003 & $ -0.029 \pm0.032 \pm$0.002 &  0.999 &  0.002 \\
 0.667 &  7.53 & $ -0.023 \pm0.020 \pm$0.001 &  0.997 &  0.005 & $ -0.019 \pm0.041 \pm$0.001 &  0.998 &  0.004 \\
 0.750 &  7.66 & $ -0.053 \pm0.026 \pm$0.003 &  0.994 &  0.002 & $  0.040 \pm0.053 \pm$0.002 &  0.995 &  0.002 \\
 0.844 &  7.78 & $  0.046 \pm0.042 \pm$0.002 &  0.980 &  0.003 & $ -0.127 \pm0.074 \pm$0.008 &  0.983 &  0.003 \\
 \hline
\multicolumn{6}{c}{$\theta\approx 10.5^\circ$; E=32.3 GeV   } \\
\hline
 0.129 &  6.09 & $ -0.014 \pm0.025 \pm$0.001 &  0.962 & -0.005 & $  0.010 \pm0.038 \pm$0.001 &  0.971 & -0.002 \\
 0.144 &  6.71 & $  0.014 \pm0.019 \pm$0.001 &  0.969 & -0.006 & $ -0.005 \pm0.031 \pm$0.001 &  0.977 & -0.002 \\
 0.162 &  7.41 & $ -0.036 \pm0.016 \pm$0.002 &  0.977 & -0.005 & $ -0.048 \pm0.027 \pm$0.003 &  0.982 & -0.002 \\
 0.182 &  8.21 & $ -0.017 \pm0.015 \pm$0.002 &  0.982 & -0.005 & $ -0.005 \pm0.024 \pm$0.001 &  0.986 & -0.002 \\
 0.205 &  9.06 & $ -0.024 \pm0.015 \pm$0.002 &  0.986 & -0.005 & $  0.027 \pm0.025 \pm$0.002 &  0.989 & -0.002 \\
 0.230 &  9.93 & $ -0.012 \pm0.016 \pm$0.001 &  0.989 & -0.005 & $ -0.024 \pm0.026 \pm$0.002 &  0.991 & -0.002 \\
 0.259 & 10.77 & $ -0.005 \pm0.017 \pm$0.001 &  0.992 & -0.005 & $ -0.010 \pm0.029 \pm$0.001 &  0.993 & -0.002 \\
 0.292 & 11.70 & $  0.026 \pm0.018 \pm$0.002 &  0.994 & -0.004 & $  0.009 \pm0.032 \pm$0.001 &  0.995 & -0.002 \\
 0.328 & 12.64 & $ -0.011 \pm0.020 \pm$0.002 &  0.995 & -0.004 & $ -0.008 \pm0.036 \pm$0.002 &  0.996 & -0.002 \\
 0.369 & 13.61 & $  0.036 \pm0.023 \pm$0.003 &  0.996 & -0.004 & $  0.043 \pm0.042 \pm$0.003 &  0.996 & -0.002 \\
 0.415 & 14.55 & $  0.016 \pm0.026 \pm$0.002 &  0.997 & -0.003 & $  0.060 \pm0.050 \pm$0.004 &  0.997 & -0.002 \\
 0.467 & 15.56 & $ -0.015 \pm0.031 \pm$0.003 &  0.998 & -0.003 & $  0.009 \pm0.062 \pm$0.003 &  0.998 & -0.002 \\
 0.526 & 16.56 & $  0.071 \pm0.038 \pm$0.004 &  0.998 & -0.002 & $ -0.125 \pm0.081 \pm$0.008 &  0.999 & -0.001 \\
 0.591 & 17.55 & $  0.045 \pm0.050 \pm$0.002 &  0.999 & -0.002 & $  0.023 \pm0.112 \pm$0.001 &  0.999 & -0.002 \\
 0.666 & 18.49 & $ -0.007 \pm0.070 \pm$0.000 &  0.999 & -0.002 & $ -0.188 \pm0.161 \pm$0.012 &  0.999 & -0.002 \\
 0.749 & 19.45 & $  0.094 \pm0.108 \pm$0.005 &  0.999 & -0.002 & $  0.338 \pm0.251 \pm$0.021 &  0.999 & -0.002 \\
 0.844 & 20.45 & $  0.309 \pm0.211 \pm$0.016 &  0.998 & -0.002 & $  0.240 \pm0.411 \pm$0.015 &  0.999 & -0.003 \\
\end{tabular}
 \end{table}
\renewcommand{\baselinestretch}{1} 
\small\normalsize	      

%% file: A2-XG2-PLB.tex

 \begin{table} 
 \caption{ Results for $A_2$ and $xg_2$  with statistical errors for proton and deuteron 
  at the measured $x$ and $Q^2$ [(GeV/c)$^2$]. The systematic error on $xg_2$ is given by $a + bx$
 where $a_p(a_d)$= 0.0016( 0.0009)  and $b_p(b_d)$=$-0.0012$($-0.0009$).  } 
\renewcommand{\baselinestretch}{.9}
\small\normalsize                   
\label{tb:A2-XG2}
 \begin{tabular}{rrrrrr}
$x$ & $<Q^2>$ &$A_2^p$ & $xg_2^p$ & $A_2^d$ & $xg_2^d$ \\
 \hline
\multicolumn{6}{c}{$\theta\approx 2.75^\circ$; E=29.1 GeV} \\
\hline
 0.021 &  0.80 & $ -0.015 \pm$ 0.012 & $ -0.037 \pm$ 0.026 & $  0.003 \pm$ 0.017 & $  0.009 \pm$ 0.036 \\
 0.026 &  0.90 & $ -0.009 \pm$ 0.008 & $ -0.026 \pm$ 0.015 & $  0.010 \pm$ 0.011 & $  0.020 \pm$ 0.021 \\
 0.038 &  1.10 & $  0.016 \pm$ 0.006 & $  0.020 \pm$ 0.010 & $ -0.013 \pm$ 0.009 & $ -0.021 \pm$ 0.014 \\
 0.061 &  1.30 & $  0.026 \pm$ 0.008 & $  0.017 \pm$ 0.009 & $ -0.017 \pm$ 0.011 & $ -0.024 \pm$ 0.013 \\
 0.098 &  1.60 & $  0.014 \pm$ 0.010 & $ -0.011 \pm$ 0.009 & $  0.025 \pm$ 0.015 & $  0.016 \pm$ 0.013 \\
 0.155 &  1.80 & $  0.061 \pm$ 0.015 & $  0.005 \pm$ 0.010 & $  0.008 \pm$ 0.024 & $ -0.005 \pm$ 0.013 \\
 0.245 &  2.00 & $  0.098 \pm$ 0.024 & $ -0.005 \pm$ 0.010 & $  0.058 \pm$ 0.038 & $  0.002 \pm$ 0.014 \\
 0.380 &  2.10 & $  0.258 \pm$ 0.064 & $  0.007 \pm$ 0.018 & $ -0.008 \pm$ 0.105 & $ -0.031 \pm$ 0.024 \\
 \hline
\multicolumn{6}{c}{$\theta\approx 5.5^\circ$; E=29.1 GeV } \\
\hline
 0.061 &  2.70 & $  0.033 \pm$ 0.036 & $  0.045 \pm$ 0.061 & $  0.059 \pm$ 0.052 & $  0.094 \pm$ 0.084 \\
 0.098 &  3.50 & $  0.029 \pm$ 0.009 & $  0.019 \pm$ 0.013 & $  0.000 \pm$ 0.014 & $ -0.009 \pm$ 0.018 \\
 0.155 &  4.40 & $  0.020 \pm$ 0.008 & $ -0.017 \pm$ 0.009 & $  0.024 \pm$ 0.012 & $  0.012 \pm$ 0.012 \\
 0.245 &  5.30 & $  0.042 \pm$ 0.011 & $ -0.021 \pm$ 0.008 & $  0.037 \pm$ 0.017 & $  0.000 \pm$ 0.011 \\
 0.380 &  6.10 & $  0.035 \pm$ 0.019 & $ -0.043 \pm$ 0.007 & $  0.086 \pm$ 0.032 & $  0.002 \pm$ 0.010 \\
 0.580 &  6.70 & $  0.107 \pm$ 0.045 & $ -0.020 \pm$ 0.006 & $  0.137 \pm$ 0.082 & $ -0.004 \pm$ 0.008 \\
 0.780 &  7.00 & $ -0.131 \pm$ 0.130 & $ -0.012 \pm$ 0.003 & $  0.444 \pm$ 0.232 & $  0.003 \pm$ 0.004 \\
 \hline
\multicolumn{6}{c}{$\theta\approx 10.5^\circ$; E=29.1 GeV} \\
\hline
 0.155 &  7.10 & $  0.030 \pm$ 0.018 & $ -0.001 \pm$ 0.024 & $ -0.023 \pm$ 0.032 & $ -0.042 \pm$ 0.039 \\
 0.245 &  9.90 & $  0.018 \pm$ 0.016 & $ -0.036 \pm$ 0.016 & $  0.029 \pm$ 0.031 & $  0.006 \pm$ 0.025 \\
 0.380 & 13.10 & $  0.054 \pm$ 0.025 & $ -0.026 \pm$ 0.013 & $  0.035 \pm$ 0.052 & $ -0.006 \pm$ 0.021 \\
 0.580 & 16.30 & $  0.090 \pm$ 0.068 & $ -0.010 \pm$ 0.010 & $  0.031 \pm$ 0.156 & $ -0.009 \pm$ 0.017 \\
 0.780 & 18.40 & $ -0.182 \pm$ 0.259 & $ -0.008 \pm$ 0.005 & $  0.795 \pm$ 0.625 & $  0.010 \pm$ 0.009 \\
 \hline
\multicolumn{6}{c}{$\theta\approx 2.75^\circ$; E=32.3 GeV} \\
\hline
 0.021 &  0.80 & $ -0.001 \pm$ 0.008 & $ -0.007 \pm$ 0.020 & $  0.003 \pm$ 0.014 & $  0.006 \pm$ 0.031 \\
 0.026 &  0.90 & $  0.002 \pm$ 0.006 & $ -0.004 \pm$ 0.014 & $ -0.006 \pm$ 0.011 & $ -0.010 \pm$ 0.022 \\
 0.038 &  1.10 & $  0.007 \pm$ 0.005 & $  0.001 \pm$ 0.009 & $  0.003 \pm$ 0.008 & $  0.006 \pm$ 0.014 \\
 0.061 &  1.30 & $  0.019 \pm$ 0.006 & $  0.009 \pm$ 0.008 & $  0.015 \pm$ 0.010 & $  0.015 \pm$ 0.013 \\
 0.098 &  1.60 & $  0.021 \pm$ 0.009 & $ -0.004 \pm$ 0.008 & $ -0.004 \pm$ 0.014 & $ -0.011 \pm$ 0.013 \\
 0.155 &  1.80 & $  0.045 \pm$ 0.013 & $ -0.008 \pm$ 0.009 & $  0.045 \pm$ 0.021 & $  0.015 \pm$ 0.013 \\
 0.245 &  2.00 & $  0.076 \pm$ 0.020 & $ -0.018 \pm$ 0.009 & $  0.063 \pm$ 0.034 & $  0.003 \pm$ 0.014 \\
 0.380 &  2.10 & $  0.209 \pm$ 0.053 & $ -0.004 \pm$ 0.017 & $  0.076 \pm$ 0.095 & $ -0.011 \pm$ 0.025 \\
 \hline
\multicolumn{6}{c}{$\theta\approx 5.5^\circ$; E=32.3 GeV } \\
\hline
 0.061 &  2.70 & $ -0.015 \pm$ 0.023 & $ -0.041 \pm$ 0.042 & $  0.046 \pm$ 0.035 & $  0.077 \pm$ 0.061 \\
 0.098 &  3.50 & $  0.017 \pm$ 0.007 & $  0.000 \pm$ 0.011 & $  0.004 \pm$ 0.011 & $ -0.003 \pm$ 0.016 \\
 0.155 &  4.40 & $  0.033 \pm$ 0.007 & $ -0.002 \pm$ 0.008 & $  0.028 \pm$ 0.010 & $  0.015 \pm$ 0.011 \\
 0.245 &  5.30 & $  0.041 \pm$ 0.009 & $ -0.023 \pm$ 0.007 & $  0.034 \pm$ 0.015 & $  0.003 \pm$ 0.010 \\
 0.380 &  6.10 & $  0.069 \pm$ 0.016 & $ -0.029 \pm$ 0.007 & $  0.000 \pm$ 0.028 & $ -0.024 \pm$ 0.009 \\
 0.580 &  6.70 & $  0.126 \pm$ 0.038 & $ -0.016 \pm$ 0.005 & $  0.078 \pm$ 0.074 & $ -0.008 \pm$ 0.007 \\
 0.780 &  7.00 & $  0.177 \pm$ 0.110 & $ -0.004 \pm$ 0.003 & $  0.170 \pm$ 0.210 & $ -0.002 \pm$ 0.004 \\
 \hline
\multicolumn{6}{c}{$\theta\approx 10.5^\circ$; E=32.3 GeV} \\
\hline
 0.155 &  7.10 & $  0.027 \pm$ 0.013 & $  0.001 \pm$ 0.019 & $  0.025 \pm$ 0.022 & $  0.019 \pm$ 0.029 \\
 0.245 &  9.90 & $  0.026 \pm$ 0.012 & $ -0.029 \pm$ 0.013 & $  0.006 \pm$ 0.021 & $ -0.016 \pm$ 0.018 \\
 0.380 & 13.10 & $  0.033 \pm$ 0.019 & $ -0.034 \pm$ 0.010 & $ -0.010 \pm$ 0.035 & $ -0.028 \pm$ 0.015 \\
 0.580 & 16.30 & $  0.000 \pm$ 0.048 & $ -0.024 \pm$ 0.008 & $  0.215 \pm$ 0.105 & $  0.013 \pm$ 0.012 \\
 0.780 & 18.40 & $ -0.146 \pm$ 0.191 & $ -0.008 \pm$ 0.004 & $ -0.527 \pm$ 0.424 & $ -0.011 \pm$ 0.007 \\
 \hline
\multicolumn{6}{c}{AVERAGE                                  } \\
\hline
 0.021 &  0.80 & $ -0.005 \pm$ 0.007 & $ -0.018 \pm$ 0.016 & $  0.003 \pm$ 0.011 & $  0.008 \pm$ 0.023 \\
 0.026 &  0.90 & $ -0.003 \pm$ 0.005 & $ -0.014 \pm$ 0.010 & $  0.002 \pm$ 0.008 & $  0.006 \pm$ 0.015 \\
 0.038 &  1.10 & $  0.011 \pm$ 0.004 & $  0.010 \pm$ 0.007 & $ -0.004 \pm$ 0.006 & $ -0.007 \pm$ 0.010 \\
 0.061 &  1.40 & $  0.020 \pm$ 0.005 & $  0.011 \pm$ 0.006 & $  0.003 \pm$ 0.007 & $ -0.001 \pm$ 0.009 \\
 0.098 &  2.30 & $  0.023 \pm$ 0.004 & $ -0.003 \pm$ 0.005 & $  0.006 \pm$ 0.007 & $ -0.001 \pm$ 0.007 \\
 0.155 &  3.70 & $  0.036 \pm$ 0.004 & $ -0.007 \pm$ 0.004 & $  0.026 \pm$ 0.007 & $  0.009 \pm$ 0.006 \\
 0.245 &  5.00 & $  0.048 \pm$ 0.005 & $ -0.022 \pm$ 0.004 & $  0.036 \pm$ 0.009 & $  0.000 \pm$ 0.005 \\
 0.380 &  7.10 & $  0.064 \pm$ 0.009 & $ -0.031 \pm$ 0.004 & $  0.029 \pm$ 0.017 & $ -0.015 \pm$ 0.005 \\
 0.580 &  8.40 & $  0.092 \pm$ 0.023 & $ -0.018 \pm$ 0.003 & $  0.122 \pm$ 0.047 & $ -0.004 \pm$ 0.005 \\
 0.780 &  8.20 & $  0.004 \pm$ 0.074 & $ -0.007 \pm$ 0.002 & $  0.228 \pm$ 0.142 & $  0.000 \pm$ 0.002 \\
\end{tabular}
 \end{table}
\renewcommand{\baselinestretch}{1} 
\small\normalsize	      

%% file: g2_lanl.bbl
\begin{thebibliography}{99}

\bibitem[\dag\dag]{W&M} Present Address: University of Massachusetts, 
Amherst, MA 01003

\bibitem[\S\S]{BNL}
Present Address: Brookhaven National Laboratory, Upton, NY 11973

\bibitem[\infty]{Fermi}
Present Address: Fermi National Accelerator Laboratory, Batavia, IL 60510

\bibitem[\flat]{Bochum}
Present Address: Ruhr-Universit${\ddot {\rm a}}$t Bochum,  Bochum, Germany

\bibitem[\S]{LANL}
Present Address: Los Alamos National Laboratory, Los Alamos, NM 87545

\bibitem[\heartsuit]{Argonne}
Present Address: Argonne National Laboratory, Argonne, IL 60439

\bibitem[\spadesuit]{Saclay}
Present Address:DAPNIA-Service de Physique Nucleaire, CEA-Saclay,
F-91191 Gif/Yvette Cedex, France 

\bibitem[\dag]{SLAC}
Present Address: Stanford Linear Accelerator Center, Stanford, CA 94305

\bibitem[\ddag]{Maryland}
Present Address: U. of Maryland, College Park, MD 20742


\bibitem{Vain} E.\ Shuryak and A.\ Vainshtein, Nuc.\ Phys.\
B {\bf 201}, 141 (1982). 
\bibitem{Jaffe}R.\ Jaffe and X.\ Ji, Phys.\ Rev.\ D {\bf 43}, 724 (1991).
\bibitem{CPR} J.\ L.\ Cortes, B.\ Pire and J.\ P.\ Ralston, Z. Phys.
C {\bf 55}, 409 (1992).
\bibitem{g2ww}S.\ Wandzura and F.\ Wilczek, Phys.\ Lett.\ B {\bf 72}, 195 (1977).
\bibitem{SMC} SMC: B.\ Adeva {\em{et al.}}, 
Phys. Rev. D {\bf 58}, 112001 (1998).
\bibitem{E143}E143: 
K.\ Abe {\em{et al.}}, Phys. Rev. D {\bf 58}, 112003 (1998).
\bibitem{E142_2}E142: P.\ Anthony {\em{et al.}}
Phys.\ Rev.\ D {\bf 54},  6620 (1996).
\bibitem{E154}E154: 
K.\ Abe {\em{et al.}}, Phys. Rev. Lett. {\bf 79}, 26 (1997).
\bibitem{E155}E155:  P. Anthony {\em{et al.}}, 
 Phys.\ Lett.\ B {\bf 463}, 339 (1999);\ B {\bf 493}, 19 (2000).
\bibitem{Hermes}HERMES: 
K. \ Ackerstaff {\em{et al.}} Phys. Lett. B {\bf 404}, 383 (1997);
A. Airapetian {\em{et al.}}, Phys. Lett.  B {\bf 442}, 484 (1998).
\bibitem{Song}X.\ Song, Phys.\ Rev.\ D {\bf 54},  1955 (1996).
\bibitem{SMCg2}SMC: 
D.\ Adams {\em{et al.}}, Phys. Lett. B {\bf 336}, 125 (1994);
{\bf 396}, 338 (1997).
\bibitem{E143_2}E143:
 K.\ Abe {\em{et al.}}, Phys. Rev.  Lett. {\bf 76}, 587 (1996).
\bibitem{E155g2} E155: P.\ Anthony  {\em{et al.}},
 Phys.\ Lett.\ B {\bf 458}, 529 (1999).
\bibitem{E154_2}E154: 
K.\ Abe {\em{et al.}}, Phys.  Lett. B {\bf 404}, 377 (1997).
\bibitem{LiD} S. \ B\"{u}ltmann  {\em{et al.}}, Nucl. Inst. Meth.  A {\bf 425},
23 (1999).
\bibitem{NMC}NMC: 
M.\ Arneodo {\em{et al.}}, Phys.\ Lett.\ B {\bf 364}, 107 (1995).
\bibitem{R1998}E143: 
K.\ Abe {\em{et al.}}, Phys. Lett. B {\bf 452}, 194 (1999).
\bibitem{Stratmann}M.\ Stratmann, Z.\ Phys.\ C {\bf 60}, 763 (1993).
\bibitem{WGR} H.\ Weigel and  L.\ Gamberg, Nucl. Phys. A {\bf 680}, 48 (2000).
\bibitem{Waka} M.\ Wakamatsu, Phys. Lett. B {\bf 487}, 118 (2000).
\bibitem{soffer}J.\ Soffer and O.\ V.\ Teryaev, Phys.\ Lett. \ B {\bf 490}, 106 (2000). 
\bibitem{JiOsborne} X. \ Ji and J. \ Osborne, Nucl. Phys. B {\bf 608}, 235 (2001).
\bibitem{LQCD}M.\ G\"{o}ckeler {\em{et al.}}, Phys. Rev. D {\bf 63}, 074506 (2001).
\bibitem{Ji}X.\ Ji and P.\ Unrau, Phys.\ Lett.\ B {\bf 333}, 228 (1994).
\bibitem{Stein}E.\ Stein {\em{et al.}}, Phys.\ Lett.\ B {\bf 343}, 369 (1995).
\bibitem{BBK}I.\ Balitsky, V.\ Braun and A.\ Kolesnichenko, Phys.\
Lett.\ B {\bf 242}, 245 (1990); {\bf 318}, 648 (1993) (Erratum).
\bibitem{Ehrnsperger}B. Ehrnsperger and A. Schafer, Phys. Rev. D {\bf 52}, 2709 (1995).
\bibitem{Weiss}J. Balla, M.V. Polyakov, and C. Weiss, 
Nucl. Phys. B {\bf 510}, 327 (1998).
\bibitem{buco}
H.~Burkhardt and W.~N.~Cottingham,  Ann. Phys. {\bf 56}, 453 (1970).
\bibitem{Ivanov} I. \ P. \ Ivanov  {\em{et al.}}, Phys. Rep. {\bf 320}, 175 (1999).
\bibitem{ELT} A. \ V. \ Efremov, O. \ V. \ Teryaev and E.\  Leader,
Phys. Rev. D{\bf 55}, 4307 (1997).



\end{thebibliography}
